\documentclass{article}
\usepackage{amsmath}
\usepackage{amssymb}
\usepackage{graphicx}
\usepackage{dcolumn}
\usepackage{bm}

\begin{document}

\title{X-ray 
Absorption Near-Edge Structure calculations with pseudopotentials. \\
Application to $K$-edge in diamond and $\alpha$-quartz}

\author{Mathieu Taillefumier$^1$, Delphine Cabaret$^1$,
Anne-Marie Flank$^2$, Francesco Mauri$^1$ \\\\
$^1$ Laboratoire de Min\'eralogie-Cristallographie de Paris, \\
Universit\'e Pierre et Marie Curie, case 115, \\
4 place Jussieu, 75252 Paris cedex 05, France \\
$^2$ Laboratoire pour l'Utilisation du Rayonnement Electromagn\'etique,\\
Centre universitaire Paris-sud, B\^at. 209D,  91898 Orsay cedex, France}

\date{\empty}

\maketitle

\abstract
We present a reciprocal-space pseudopotential scheme 
for calculating X-ray absorption near-edge structure (XANES) 
spectra. The scheme incorporates a recursive method 
to compute absorption cross section as a continued fraction. 
The continued fraction formulation of absorption is 
advantageous in that it permits the treatment of core-hole 
interaction  through large supercells (hundreds of atoms). 
The method is compared with recently developed Bethe-Salpeter 
approach.  The method is applied to the carbon $K$-edge in 
diamond and to the silicon and oxygen $K$-edges in 
$\alpha$-quartz for which polarized XANES spectra were 
measured. Core-hole effects are investigated by varying 
the size of the supercell, thus leading to information 
similar to that obtained from cluster size analysis usually 
performed within multiple scattering calculations.



\section{Introduction}

XANES spectroscopy is a powerful technique to  probe the 
empty states in solids. Moreover XANES gives information about
the atomic arrangement around the probed atom up to the
medium range order (i.e., about 8~\AA). Nevertheless the 
interpretation of XANES spectra is not straightforward and 
often requires sophisticated simulation tools.  To calculate 
a XANES spectrum, one needs to compute the absorption cross
section. This is given by the Fermi golden rule as a
sum of probabilities per unit of time of making a transition
from an initial state to an unoccupied final state through an 
interaction Hamiltonian \cite{Brouder90}. Electronic transitions 
involved in X-ray absorption spectroscopy are mainly governed 
by the electric dipole operator. The main difficulty of any 
absorption cross section calculation lies in the solution of 
the Schr\"odinger equation for the final and initial states.
The choice of the method used to solve the Schr\"odinger equation
depends on the localized or delocalized character of final states.
Crystal field multiplet theory \cite{Cowan81} is adapted to 
the calculation of localized final states, i.e., with strong 
electron-electron interactions. It is typically the case of 
$L_{2,3}$ edges of transition elements \cite{DeGroot94} and
$M_{4,5}$ edges of rare earth elements \cite{Finazzi95}. 
If the final state is delocalized (the case of $K$ and 
$L_1$ edges) multielectronic interactions are weak and 
monoelectronic approaches based on the density functional 
theory (DFT) are usually employed. Among monoelectronic methods, 
one distinguishes the real space (cluster) approach from the 
reciprocal space (band structure) approach. Real space 
multiple scattering theory \cite{Natoli80,Fonda92,Ankudinov98} 
has been extensively used in the past twenty years. However, 
multiple scattering theory suffers from an unavoidable drawback, 
i.e., the ``muffin-tin'' approximation of the potential. Recently, 
the ``muffin-tin'' approximation has been overtaken in a real 
space approach by using the finite difference method (MDF) to solve 
the Schr\"odinger equation \cite{Joly01}. Although it yields  good
results \cite{Joly99}, the method presented in Ref.~\cite{Joly01} 
requires significant computing power. At present this is what restricts 
the applications to small clusters (around 50 atoms without any symmetry,
at maximum).  

With regards to band structure calculations, local projected densities 
of empty states (commonly called LDOS) have been often used
to interpret XANES spectra (see for instance 
Refs.~\cite{Weng89-2,Czyzyk92,Bacewicz00}). Recently several band 
structure codes have incorporated the radial matrix elements 
\cite{Muller-bis} to properly calculate XANES or ELNES (electron 
energy-loss near-edge structure). It is notably the case of
\verb!Wien2k! \cite{Wien2k}, which is based on the full-potential 
linearized augmented plane wave method \cite{Nelhiebel99},
the orthogonalized-linear-combination-of-atomic-orbital 
program developed by Mo and Ching \cite{Mo00,Mo01} and 
\verb!CASTEP!, whose ELNES part has been developed by Pickard 
\cite{Pickard97,Jayawardane01,Suenaga01}. This last approach uses 
pseudopotentials and reconstructs {\it all electron} wave functions 
within the projector augmented wave (PAW) method of Bl\"ochl 
\cite{Blochl94}. Within these {\it ab initio} schemes the
treatment of core hole-electron interactions is usually achieved 
through the inclusion of an excited atom into a supercell. 
The absorbing atom is then considered as an impurity. Beside these 
methods, Shirley \cite{Shirley98} and more recently Soininen and 
Shirley \cite{Soininen01} have proposed alternative formalisms 
that treat core-hole interactions in a different way. Indeed their 
methods incorporate core-hole interactions in the two-particle 
Bethe-Salpeter (BS) equation. Such methods are of great importance 
for optical absorption calculations 
\cite{Albrecht98,Benedict98-a,Benedict98-b,Rohlfing98-a}. However,
in the case of $K$-edge XANES spectra, the core-hole is frozen
at one atomic site and the two-particle BS approach can be reduced to
a single particle calculation.

By using supercell approaches, large systems (several hundreds of atoms)
can be {\it a priori} treated. The charge density for such large cells 
is easy to calculate self-consistently by using pseudopotentials. 
On the other hand, the computation of the cross section is at present 
limited by the diagonalisation of the Hamiltonian for several empty 
states at many $\bm{k}$-points in the Brillouin zone.  Indeed in 
Refs.\cite{Pickard97,Jayawardane01} pseudopotential calculations 
of ELNES spectra in diamond and $c$-BN have been performed with a small 
16-atom supercell. Here we propose to use the recursion method of
Haydock, Heine and Kelly \cite{Haydock72,Haydock75,Haydock80} to make 
the calculation of the cross section computationally tractable for 
larger supercells. Thanks to this iterative technique, the calculation 
time of XANES becomes negligible compared with the calculation of 
the self-consistent charge density in large systems. 

We incorporated the recursion method into a pseudopotential scheme. 
We first show in section 2 that the absorption cross section can be 
easily computed within a pseudopotential formalism using the PAW method. 
Then, by using the recursion method, we express the absorption cross 
section as a continued fraction. We present applications to $K$-edges 
in section 3. Two materials have been chosen : diamond which is an 
interesting example for method testing, and $\alpha$-quartz for which 
we have measured polarized XANES spectra.  

\section{Method}

\subsection{X-ray absorption cross section in the impurity model}

In a monoelectronic approach, the X-ray absorption cross section 
$\sigma(\omega)$ can be written  as

\begin{equation}
\sigma(\omega)=4\pi^2\alpha_0 \hbar \omega \sum_f |M_{i\rightarrow f}|^2
\delta(E_f-E_i-\hbar\omega),
\end{equation}
where $\hbar\omega$ is the photon energy, 
$\alpha_0$ is the fine structure constant and $M_{i\rightarrow f}$ 
are the transition
amplitudes between an initial core state $|\psi_i\rangle$ with energy $E_i$,
localized on the absorbing atom site ${\bm R_0}$,
and an {\it all electron} final
state $|\psi_f\rangle$ with energy $E_f$
 
\begin{equation}
M_{i\rightarrow f}=\langle\psi_f |\mathfrak{O}| \psi_i \rangle.
\label{matrix_elt}
\end{equation} 
$\mathfrak{O}$ is a transition operator coupling
initial and final states. In the electric quadrupole 
approximation, $\mathfrak{O}$ is given by
$\hat{\bm{\varepsilon}}\cdot{\bm r}(1+\frac{i}{2}\bm{k}\cdot\bm{r})$, 
where $\hat{\bm{\varepsilon}}$ and $\bm{k}$ 
are  the polarisation vector and the wave vector of the photon 
beam, respectively. Within the frozen core approximation, 
$|\psi_i\rangle$ is a core state that can be taken from an {\it all electron} 
ground state atomic calculation. In the impurity model, $|\psi_f\rangle$
is an excited empty state that is solution of the Schr\"odinger equation 
for a potential that includes a core-hole on the absorbing atom. 

\subsection{PAW formalism}

In the following we want to express the transition amplitude 
(Eq.\ref{matrix_elt}) within the PAW formalism, as originally 
described by Bl\"ochl \cite{Blochl94}. We only recall the main 
aspects of the method that are needed to give a simple 
expression for the $M_{i\rightarrow f}$ terms.

Within the PAW formalism, the final state {\it all electron} 
wave functions $|\psi_f\rangle$ are related to the corresponding final 
pseudo wave functions $|\widetilde{\psi}_f\rangle$ through a linear
operator $\mathcal{T}$:
\begin{equation}
|\psi_f\rangle = \mathcal{T}|\widetilde{\psi}_f\rangle.
\label{psif}
\end{equation}
$\mathcal{T}$ differs from identity by a sum of local
atom-centered contributions, that act only within spherical core regions
centered on each atomic site ${\bm R}$, called  augmentation regions
or $\Omega_{\bm R}$:
\begin{equation}
\mathcal{T}={\bf 1} + \sum_{{\bm R},n}\left[|\phi_{{\bm R},n}\rangle
-|\widetilde{\phi}_{{\bm R},n}\rangle\right]\langle\widetilde{p}_{{\bm R},n}|.
\label{T}
\end{equation}
Here $|{\phi}_{{\bm R},n}\rangle$ and $|\widetilde{\phi}_{{\bm R},n}\rangle$
are the {\it all electron} and pseudo partial waves, 
respectively, which coincide outside 
$\Omega_{\bm R}$. The vectors $\langle\widetilde{p}_{{\bm R},n}|$,
called projector functions \cite{Blochl94},  are equal 
to zero outside $\Omega_{\bm R}$ and satisfy the condition
$\langle \widetilde{p}_{{\bm R},n}|
\widetilde{\phi}_{{\bm R}^\prime,n^\prime}\rangle=
\delta_{{\bm{RR}}^\prime}\delta_{nn^\prime}$.
The index $n$ refers to the angular momentum quantum numbers
$(\ell,m)$ and to an additional number, used if there is more 
than one projector per angular momentum channel.
The $|{\phi}_{{\bm R},n}\rangle$ form a complete basis for 
any physical non-core {\it all electron} wave function within 
$\Omega_{\bm R}$\cite{footnote}. Therefore the 
$|\widetilde{\phi}_{{\bm R},n}\rangle$
are also a complete basis for any physical pseudo 
wave function $|\widetilde{\psi}\rangle$ within $\Omega_{\bm R}$, i.e. 
for any $\langle \bm{r}|\chi_{\bm{R}}\rangle$ function 
centered on an atomic site $\bm{R}$ and equal to zero outside 
$\Omega_{\bm R}$, 
\begin{equation}
\sum_n \langle \widetilde{\psi}|\widetilde{p}_{{\bm R},n}\rangle
\langle\widetilde{\phi}_{{\bm R},n}|\chi_{\bm{R}}\rangle
= \langle\widetilde{\psi}|\chi_{\bm{R}}\rangle.
\label{condition}
\end{equation}

Substituting Eq. (\ref{T}) in Eq. (\ref{psif}) and then Eq. (\ref{psif})
in Eq. (\ref{matrix_elt}), 
the transition amplitude $M_{i\rightarrow f}$ becomes
\begin{eqnarray}
M_{i\rightarrow f}&=&\langle\widetilde{\psi}_f |\mathfrak{O}|\psi_i\rangle
\nonumber \\
& & +\sum_{{\bm R},n}\langle\widetilde{\psi}_f |
    \widetilde{p}_{{\bm R},n}\rangle
    \langle \phi_{{\bm R},n}|\mathfrak{O}| \psi_i \rangle
    \nonumber \\
& & -\sum_{{\bm R},n} \langle\widetilde{\psi}_f |
    \widetilde{p}_{{\bm R},n}\rangle
    \langle \widetilde{\phi}_{{\bm R},n}|\mathfrak{O}| \psi_i \rangle.
\label{general}
\end{eqnarray}
In Eq.\ref{general}, the initial wave function 
$\langle\bm{r}|\psi_i\rangle$ is localized on the site 
of the absorbing atom, $\bm{R_0}$, then only the $\bm{R_0}$ term
has to be considered in each sum. Furthemore it should be noticed that 
$\langle \bm{r}|\mathfrak{O}|\psi_i\rangle$
is zero outside the $\Omega_{\bm R_0}$ region. Therefore  
we can make use of Eq.(\ref{condition}) 
for the third term of Eq.(\ref{general}), which thus vanishes with
the first term. The transition amplitude $M_{i\rightarrow f}$
is then reduced to one term. Now, introducing
\begin{equation}
|\widetilde{\varphi}_{\bm R_0}\rangle = \sum_n 
|\widetilde{p}_{{\bm R_0},n}\rangle
\langle \phi_{{\bm R_0},n}|
\mathfrak{O}| \psi_i \rangle,
\label{phi}
\end{equation}
we obtain the following simple expression for the X-ray absorption 
cross section

\begin{equation}
\sigma(\omega)=4\pi^2\alpha_0\hbar\omega \sum_f |\langle\widetilde{\psi}_f|
\widetilde{\varphi}_{\bm R_0}\rangle|^2
\delta(E_f-E_i-\hbar\omega).
\label{sigma}
\end{equation}

The calculation of XANES spectra from Eq.\ref{sigma} presents 
us with the problem of determining many empty states 
$|\widetilde{\psi}_f\rangle$, as has been mentioned in 
section 1. Indeed, the addition of unoccupied bands
significantly increases computing time and then limits
the size of the supercells. In the following, we show how
the recursion method  permits the cross section (Eq.\ref{sigma}) 
to be rewritten as a continued fraction, so that
only occupied bands have to be calculated.
 
\subsection{Recursion method}

The recursion method comprises of a powerful recursive algorithm that can 
be applied to a Hermitian matrix in order to transform it into 
a tridiagonal form. This method has been widely used in solid states physics.
The pioneering works in this field were mainly by Haydock, Heine and Kelly
\cite{Haydock72,Haydock75}. The first application to absorption 
cross section calculation
was carried out by Filipponi within the multiple scattering formalism
\cite{Filipponi91}. In a close field Benedict and Shirley have 
implemented it for $\varepsilon_2(\omega)$ calculation 
including core-hole interaction \cite{Benedict99} within the BS approach.
For simplicity, in the following, we assume that the norm of {\it all
electron} partial waves coincide with that of the pseudo-partial waves.

In order to use the recursion method in 
our scheme, we shall introduce
the Green operator into Eq.(\ref{sigma}) by making the substitution

\begin{equation}
\sum_f|\widetilde{\psi}_f\rangle \delta(E_f-E_i-\hbar\omega)
\langle \widetilde{\psi}_f|= - \frac{1}{\pi}\mathfrak{Im}[\widetilde{G}(E)],
\end{equation}
with 
\begin{equation}
\widetilde{G}(E)=(E - \widetilde{H} + i \gamma)^{-1}.
\label{Green}
\end{equation}
In Eq.(\ref{Green}) $\widetilde{G}(E)$ is the Green operator associated 
with the pseudo-Hamiltonian $\widetilde{H}=\mathcal{T}^\dagger H\mathcal{T}$, 
which is Hermitian.
The energy $E$ is given by
$E=E_i+ \hbar\omega $ and $\gamma$ is an infinitesimal positive 
number.
The cross section (Eq.\ref{sigma}) can be rewritten as
\begin{equation}
\sigma(\omega)=-4\pi\alpha_0\hbar\omega
\mathfrak{Im}\left[\langle\widetilde{\varphi}_{\bm{R_0}}|(E - \widetilde{H} 
+ i \gamma)^{-1}| \widetilde{\varphi}_{\bm{R_0}}\rangle\right].
\label{sigma2}
\end{equation}

Following the original work of Lanczos \cite{Lanczos50,Lanczos52},
the recursion method sets up a new basis in which 
the pseudo-Hamiltonian $\widetilde{H}$,
has a tridiagonal representation, from which the matrix elements
$\langle\varphi_{\bm{R_0}}|(\widetilde{H}-E-i\gamma)^{-1}|
\varphi_{\bm{R_0}}\rangle$ are very simply
derived. This new basis (called Lanczos basis in the following) 
is obtained by the repeated action
of $\widetilde{H}$ onto the normalized initial 
vector 
$|u_0\rangle=|\widetilde{\varphi}_{\bm{R_0}}\rangle/\sqrt{
\langle\widetilde{\varphi}_{\bm{R_0}}|
\widetilde{\varphi}_{\bm{R_0}}\rangle}$
through the symmetric three-term recurrence relation 
\begin{equation}
\widetilde{H}|u_i\rangle=a_i|u_i\rangle + b_{i+1}|u_{i+1}\rangle + 
b_i|u_{i-1}\rangle.
\label{recurrence}
\end{equation}
where $\{a_i\}$ and 
$\{b_i\}$ are two sets of real parameters given by 
$a_i=\langle u_i|\widetilde{H}|u_i\rangle$ and 
$b_i=\langle u_i|\widetilde{H}|u_{i-1}\rangle=
\langle u_{i-1}|\widetilde{H}|u_i\rangle$. 
The tridiagonal matrix representation
of $\widetilde{H}$ in the $\{|u_i\rangle\}$  basis then leads to the
following form of the matrix elements of Eq.~(\ref{sigma2})
\begin{equation}
\langle\widetilde{\varphi}_{\bm{R_0}}|(\widetilde{H}-E-i\gamma)^{-1}|
\widetilde{\varphi}_{\bm{R_0}}\rangle =
\frac{\langle\widetilde{\varphi}_{\bm{R_0}}|
\widetilde{\varphi}_{\bm{R_0}}\rangle}
{a_0-E-i\gamma-\frac{b_1^2}{a_1-E-i\gamma-
\frac{b_2^2}{\ddots}}}.
\label{fraction}
\end{equation}

A simple terminator has been used to finish the continued fraction 
\cite{Benedict99}. In particular, if $N$ is the number of iterations 
required to converge the calculation, we consider that the 
coefficients $(a_i,b_i)$ are equal to $(a_N,b_N)$ for $i>N$, 
this leads to an analytical form of the terminator. It should 
be noted that the number of iterations $N$ strongly depends on 
the broadening parameter $\gamma$. With the iterative technique 
of Haydock the main part of XANES calculation involves the 
computation of the Hamiltonian acting on a single vector. This 
means that the computing time is considerably reduced  compared 
to that can be required by Eq.~\ref{sigma} with an explicit 
diagonalisation.  

\subsection{Comparison with Bethe-Salpeter approach}

Recently, the BS approach has been successfully applied to the
calculation of XANES \cite{Shirley98,Soininen01}.
Since the formalism  of such two particles excited states 
calculations has been recently detailed in a review
\cite{Onida02}, here we only recall the main aspects.
The BS procedure \cite{Soininen01} consists of
a ground state DFT calculation in order to obtain the Kohn-Sham 
orbitals, a $GW$ calculation to correct
the eigenvalues \cite{Hedin65}, and a  solution of BS equation 
using $GW$ eigenvalues and a statically  screened potential.

Our method differs from  the BS approach of Soininen and Shirley 
\cite{Soininen01} by three points. The first two differences are 
substantial, the last one is just methodological. In particular, 
(i) we use the DFT eigenvalues, whereas Ref. \cite{Soininen01}
uses $GW$ corrected eigenvalues \cite{Hedin65}. If the $GW$ correction 
on the empty states can be described by a rigid shift, as it is often 
the case, this difference will not affect the XANES spectra.
(ii) We screen the core-hole by the valence electron response at 
all orders computed self-consistently within DFT. In 
Ref. \cite{Soininen01}, the core-hole is screened at the 
linear order using the random phase approximation \cite{Hybertsen87}.
(iii) We develop the final state wave functions $|\widetilde{\psi_f}\rangle$
in a plane wave basis set of a supercell. In Ref. \cite{Soininen01},
the BS calculation uses as basis the eigenstates of the
ground state Hamiltonian.

In the next section, our results will be compared with BS calculations
as far as possible.
 
\section{Applications}

Our method was applied to the carbon $K$-edge in 
the diamond phase and to the silicon and oxygen $K$-edges in 
$\alpha$-quartz. XANES calculations were  performed in two steps:
(i) the evaluation of the self-consistent charge density for supercells
including one $1s$ core-hole, 
(ii) the construction of a converged $\{|u_i\rangle\}$ basis 
leading to the tridiagonal representation of the pseudo-Hamiltonian.
The transition matrix elements were determined in the electric
dipole approximation. Calculations were carried out using an 
{\it ab initio} total energy code based on density functional
theory within the local density approximation (LDA)\cite{paratec}.
Norm-conserving Troullier-Martins pseudopotentials \cite{Troullier91} 
with a single component for each ($\ell,m$) component were used. 
For the excited atom a pseudopotential with only one $1s$ electron 
was generated. In both diamond and $\alpha$-quartz cases,
the size of the supercell was increased until the neighboring
excited potentials did not interact with each other. The wave functions 
were expanded in plane waves with an energy cutoff of 50~Ry and 70~Ry 
for diamond and $\alpha$-quartz respectively. Reciprocal-space 
integrations were performed using a Monkhorst-Pack ${\bm k}$-point 
grid \cite{Monkhorst76}.

\subsection{Diamond}

\begin{figure}
\includegraphics{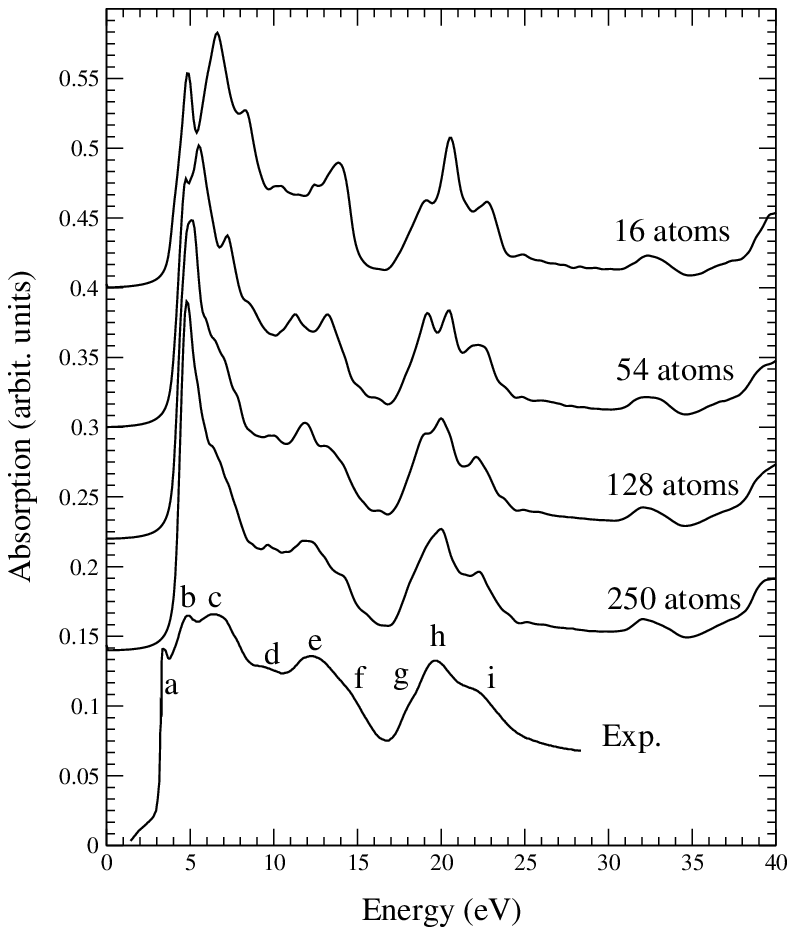}
\caption{Calculated C $K$-edge X ray absorption spectra in diamond
for different supercell sizes, compared with experimental data
(from Ref.~\cite{Ma92}). A 286.1~eV shift was applied to the
experimental spectrum.}
\label{diamond}
\end{figure}

Figure \ref{diamond} shows calculations performed with 
four different sizes of supercell, all including a $1s$
core-hole on one of the carbon atoms. The size of the 
primitive cell -- trigonal cell ($a$=2.524~\AA) 
containing two carbon atoms -- was successively multiplied 
in the three directions by a factor two 
(2$\times$2$\times$2 supercell, 16 atoms), three
(3$\times$3$\times$3 supercell, 54 atoms), four 
(4$\times$4$\times$4 supercell, 128 atoms) and five
(5$\times$5$\times$5 supercell, 250 atoms).
Further increasing the supercell size does not provide significant
modifications in the calculated spectrum. ${\bm k}$-point 
convergence of the charge density was obtained with a 
2$\times$2$\times$2 ${\bm k}$-point grid for both 2$\times$2$\times$2
and 3$\times$3$\times$3 supercells, and with only one ${\bm k}$-point
for larger supercells. ${\bm k}$-point convergence of the Lanczos basis 
was reached with a 10$\times$10$\times$10 ${\bm k}$-point grid for the 
2$\times$2$\times$2 supercell, a 8$\times$8$\times$8 ${\bm k}$-point 
grid for the 3$\times$3$\times$3 supercell, and 6$\times$6$\times$6 
${\bm k}$-point grids for both 4$\times$4$\times$4 and 5$\times$5$\times$5 
supercells. The number of iterations of the Lanczos basis was checked 
with $\gamma=$0.3~eV. The required number of $|u_i\rangle$ vectors 
varied from around 800 to 1400, while increasing the supercell size 
from 16 atoms to 250 atoms. 

Calculations are compared with the carbon $K$-XANES spectrum of diamond 
measured by Ma {\it et al.} \cite{Ma92}. The overall agreement between 
the experimental spectrum and the 5$\times$5$\times$5 supercell 
calculated one is quite good. In particular, the features from $d$ 
to $i$ are correctly reproduced as well in relative intensity as 
in energy position. Nevertheless in the first four eV of the spectrum
the agreement between experiment and calculation remains unsatisfactory. 
The intensity of peak $b$ is exagerated and the exciton peak $a$ is 
not reproduced. Looking at the various supercell  calculations, one
notices that the first 25 eV of the spectrum are affected by 
the size of the cell, i.e. the interaction between core-holes 
belonging to neighboring cells.  This gives information about the
volume seen by the photoelectron during the absorption process: 
the photoelectron interacts with atoms located within a 6~\AA\ radius 
sphere centered on the absorbing atom.

Several calculations of the carbon $K$-edge in diamond have been 
reported in the literature. Weng {\it et al.} \cite{Weng89-1} 
present pseudo-atomic-orbital density of $p$ empty states together 
with multiple scattering calculations. On the one hand, ``muffin-tin'' 
potentials as required in multiple scattering theory appear to be 
inappropriate in that case. This result was confirmed by a 300
atom cluster calculation performed by the FEFF8 code 
\cite{Ankudinov98,Rehr02}. On the other hand, although it neglects 
core-hole effects, LDOS calculation by Weng {\it et al.} 
\cite{Weng89-1} seems to correctly reproduce features $h$ and $i$.
Pickard\cite{Pickard97} shows that including core-hole effects into 
a 2$\times$2$\times$2 supercell significantly improves the agreement 
with experiment. However, as the supercell size increases, peak $b$ 
becomes much too intense (figure \ref{diamond}). This may suggest an 
over estimation of the core-hole effects in that case. This result has 
also been observed by Soininen and Shirley\cite{Soininen01} while 
core-hole interactions is treated within BS approach. Their  
carbon $K$ edge calculation in diamond, is however very similar
to our 128 and 250 atom supercell spectra. It seems that the corrections 
brought to excited states by $GW$ calculations has only a negligible 
influence on the energy positions of calculated features. This result 
is not surprising since previous studies\cite{Rohlfing93,Arnaud00} 
have shown that the $GW$ approximation used for band structure 
calculation in diamond, amounts to applying a scissors-operator shift 
which is valid to within a 0.5~eV error. This error is comparable
to the energy resolution of XANES features. The exageration of peak $b$ 
in both calculations requires more understanding and the problem of the 
treatment of core-hole interaction in diamond is at this time still unclear.

\subsection{$\alpha$-quartz}

In the case of $\alpha$-quartz, calculations are compared with
polarized Si and O $K$-XANES experiments. $\alpha$-quartz 
($\alpha$-SiO$_2$) has a trigonal space group
with a hexagonal unit cell. Since the point group is $32$ (or $D_3$),
$\alpha$-quartz is a dichroic compound \cite{Brouder90}. Therefore any 
absorption spectrum can be expressed as a linear combination
of the two spectra $\sigma_\parallel$ and $\sigma_\perp$
corresponding to the polarisation vector $\bm{\varepsilon}$ of the 
photon beam parallel and perpendicular
to the ternary axis of the hexagonal cell (i.e. the $c$ axis), respectively.
While calculations of isotropic XANES spectra of $\alpha$-quartz have been 
the subject of several studies in the past 
\cite{Jollet93,Chaboy95,Tanaka95,Wu98,Mo01}, little attention 
has been paid to the angular dependence of absorption.
Lagarde {\it et al.} \cite{Lagarde92} show that the differences
observed between $\sigma_\parallel$ and $\sigma_\perp$ spectra are 
similar to that observed between the spectra of silica and 
densified silica. In Ref. \cite{Lagarde92,Bart93} polarization 
effects are interpreted in terms of geometrical considerations.
However, to our knowledge no convincing calculations of polarized data 
have been reported yet. 
 
Si and O $K$-edge measurements were carried out respectively 
on the SA32 and SA72 beam lines at the Super-ACO facility of 
LURE (Orsay, France). The storage ring was operating at 800~MeV 
($\lambda_c = 18.6$~\AA ) and 200~mA electron current. The sample 
was a synthetic ($10\bar{1}0$) oriented single crystal
of $\alpha$-quartz. The SA32 beam line was equipped with a double 
InSb (111) crystal monochromator. The instrumental resolution was 
estimated to be around 0.7~eV. The incident beam was monitored by an 
ionisation chamber while the absorption of the sample was measured in 
the electron drain current mode. Energy calibration was checked after 
each scan by measuring the $c$-Si spectrum for which the inflection 
point of the absorption edge was set to 1839~eV. The spectra were 
collected over a photon energy range of 1830-1910~eV with 0.2~eV 
steps and 1~s integration time. The sample was placed on a rotating 
holder, the rotation axis of which was parallel to the photon beam.
The ($10\bar{1}0$) plane of the sample was set normal to the photon 
beam and the spectra were measured for the polarisation vector 
$\bm{\varepsilon}$ either parallel to the $c$ axis ($\sigma_\parallel$) 
or perpendicular to the $c$ axis ($\sigma_\perp$). We verified that 
the combination $\frac{2}{3}\sigma_\perp + \frac{1}{3} \sigma_\parallel$ 
reproduced the powder sample well. The X-ray natural linear dichroism 
(XNLD) is obtained by the difference $\sigma_\parallel - \sigma_\perp$.
The SA72 beam line was equipped with a toroidal grating monochromator. 
The estimated resolution was around 0.3~eV. Polarized O $K$-XANES were 
measured in the fluorescence mode over a photon energy range of 
525-580~eV with 0.11~eV steps. The first 12~eV were slightly affected 
by self-absorption and the signal to noise was not good enough to permit 
an unambiguous extraction of the XNLD.

Calculations were performed with various sizes of cell built from 
the cristallographic parameters given in Ref.~\cite{Will88}.
Convergence in terms of supercell size was obtained for a 
2$\times$2$\times$2 supercell (72 atoms) at both silicon and 
oxygen $K$-edges. In order to show the influence of core-hole
effects we also present results obtained from the primitive cell
(9 atoms) with or without core-hole (ground state calculation). 
Charge density was calculated with 2$\times$2$\times$2
${\bm k}$-point grid for primitive cells (with or without core-hole)
and only one ${\bm k}$-point for the 2$\times$2$\times$2 supercell.
The Lanczos basis was determined with 4$\times$4$\times$4
${\bm k}$-point grid for the primitive cell and 3$\times$3$\times$3 
${\bm k}$-point grid  for the supercell. The broadening parameter 
$\gamma$, was set to 1~eV, this led to the construction of around 
500 and 600 $|u_i\rangle$ vectors for the primitive cell and the 
supercell calculations respectively.

\begin{figure}
\includegraphics{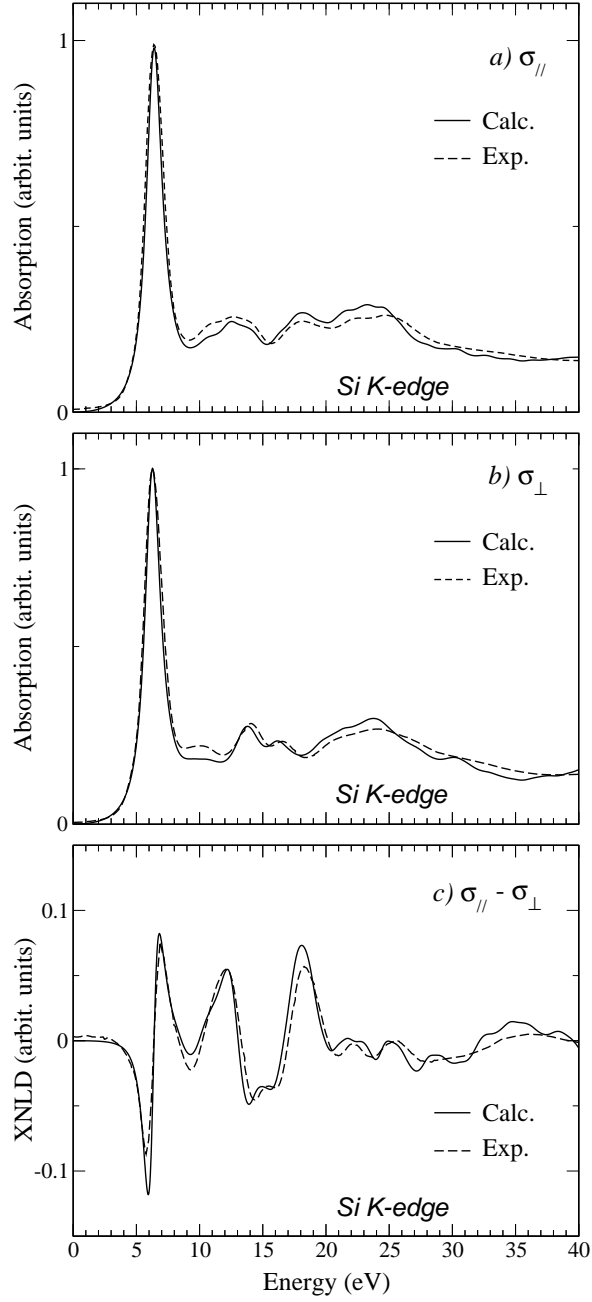}
\caption{Experimental (dashed line) and calculated (solid line) Si
$K$-edge polarized X-ray absorption spectra in $\alpha$-quartz:
(a) $\sigma_\parallel$ corresponding to $\bm{\varepsilon}\parallel[001]$;
(b) $\sigma_\perp$ corresponding to $\bm{\varepsilon}\perp[001]$;
(c) XNLD or $\sigma_\parallel$-$\sigma_\perp$.
A 1840.7~eV shift was applied to experimental data.}
\label{seuilKSi}
\end{figure}

\begin{figure}
\noindent\includegraphics[angle=-90]{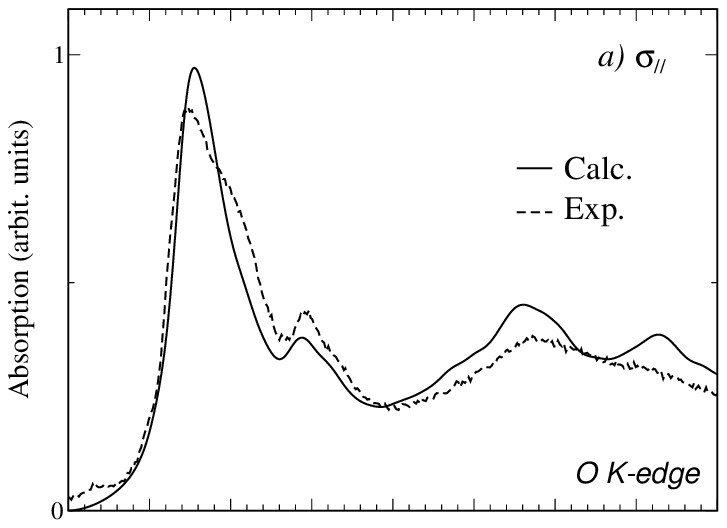}
\includegraphics[angle=-90]{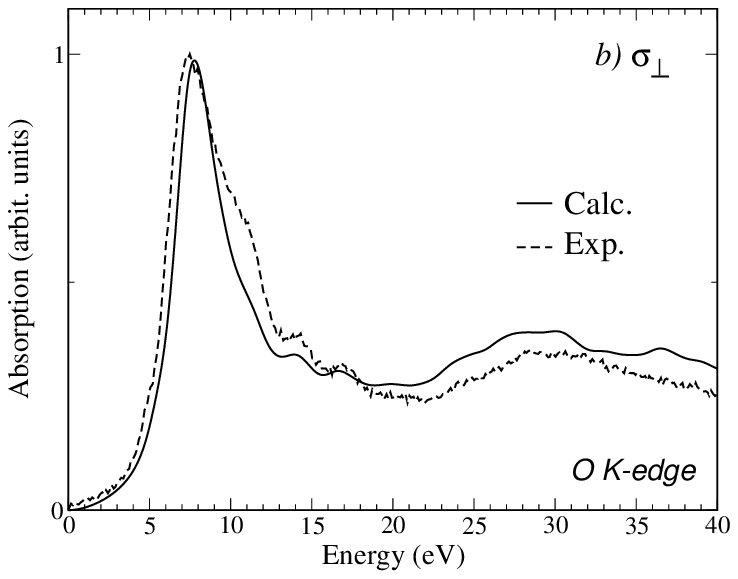}
\caption{Experimental (dashed line) and calculated (solid line) O
$K$-edge polarized X-ray absorption spectra in $\alpha$-quartz:
(a) $\sigma_\parallel$; (b) $\sigma_\perp$. A 531.9~eV shift was
applied to experimental data.}
\label{seuilKO}
\end{figure}

Experimental Si $K$-edge X-ray absorption spectra are compared with 
2$\times$2$\times$2 supercell calculations  in figure~\ref{seuilKSi}. 
From top to bottom one can see $\sigma_\parallel$ contribution 
(figure~\ref{seuilKSi}a), $\sigma_\perp$ contribution 
(figure~\ref{seuilKSi}b), and the XNLD signal (figure~\ref{seuilKSi}c). 
Figure~\ref{seuilKSi}c shows that $\alpha$-quartz presents a strong angular
dependence at the silicon $K$-edge, close to 9\% of the white line's
intensity. Experimental and calculated  spectra -- polarized and dichroic-- 
are in excellent agreement. The strength and the energy position of all 
the features are correctly reproduced. At the oxygen $K$-edge (figures 
\ref{seuilKO}a and \ref{seuilKO}b) the agreement with experiment is also 
satisfactory. $\sigma_\parallel$ and $\sigma_\perp$ spectra differ 
essentially in the 12-19~eV energy range. These differences are well 
reproduced by the calculations.

\begin{figure}
\noindent\includegraphics[angle=-90]{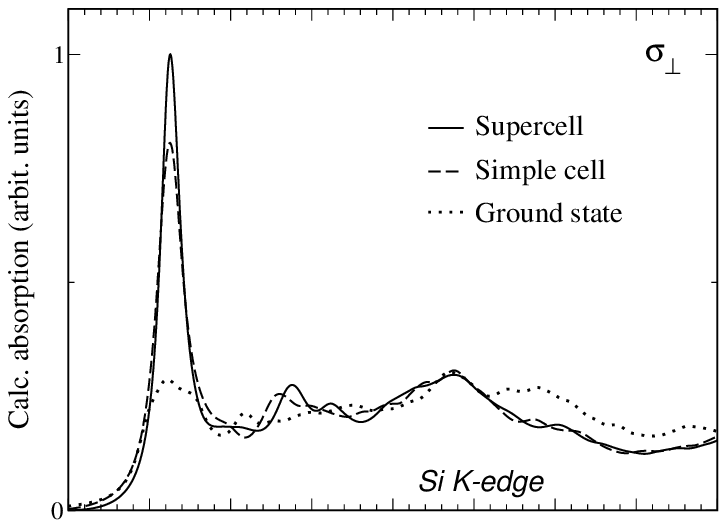}
\includegraphics[angle=-90]{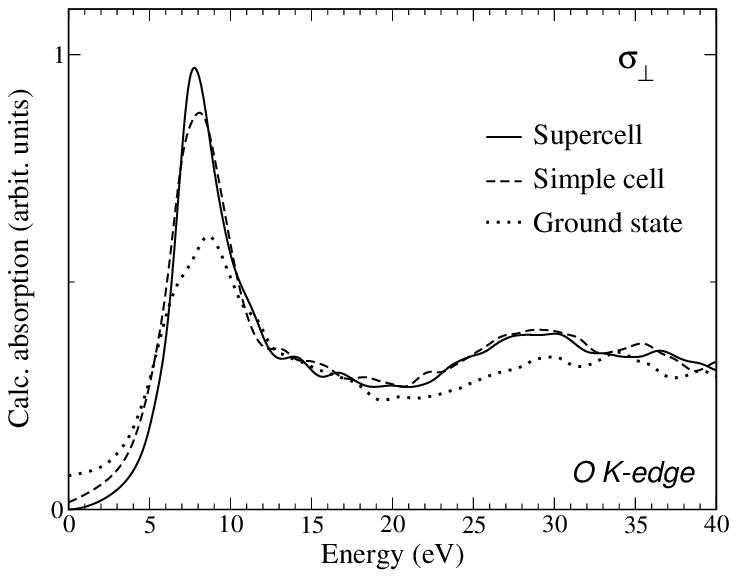}
\caption{Calculated $\sigma_\perp$ at the silicon $K$-edge (upper graph)
and at the oxygen $K$-edge (lower graph) for different cells: a
2$\times$2$\times$2 supercell including one $1s$ core-hole (solid line),
a simple cell (hexagonal unit cell) also including one $1s$ core-hole
(dashed line), and the ground state hexagonal unit cell (dotted line).}
\label{dtc}
\end{figure}

Figure \ref{dtc} compares three calculated $\sigma_\perp$ spectra 
at both silicon and oxygen $K$-edges: the 2$\times$2$\times$2 
supercell spectrum (solid line), the primitive cell calculation 
including core-hole effects (dashed line) and the ground state 
calculation (dotted line). At the silicon $K$-edge (upper panel), 
the ground state calculation does not reproduce the intense white 
line, which is clearly a signature of core hole-electron interaction.
Including a $1s$ core-hole on one of the three silicon atoms of the 
unit cell drastically modifies the spectrum and results in the 
appearance of the white line. Increasing the size of the cell 
yields changes that are essentially in the 8-18~eV energy range, 
showing that the resonances occuring in that region are related 
to medium range organisation around the absorbing atom. At the 
oxygen $K$-edge (lower panel), although core-hole effects are less
important than at the silicon $K$-edge, they definitely have to be 
taken into account in the calculation. Furthemore our simulations
show the sensitivity to medium range order of the 12-20~eV region 
just above the main peak.

The importance of core-hole effects in $\alpha$-quartz was first 
noticed by Jollet and Noguera \cite{Jollet93}. This point was 
emphasized in multiple scattering studies \cite{Chaboy95,Wu98}.
Our calculations confirm these previous considerations and 
demonstrate that reciprocal space calculations can also evidence 
the influence of medium range order, as is usually shown in 
cluster size analysis performed within the multiple scattering 
formalism \cite{Chaboy95,Wu98}.

Figures \ref{seuilKSi} and \ref{seuilKO} display a good agreement 
between calculated and experimental spectra. Nevertheless the 
agreement in energy position may be discussed to a certain extent. 
Indeed calculated features appear slightly too contracted compared  
with experiment. For example, 1~eV shift is observed at 25~eV 
in figure \ref{seuilKSi}. This contraction becomes more pronounced 
as energy increases. This small discrepancy could be due to the use 
of DFT eigenvalues. Unfortunately, in $\alpha$-quartz, the $GW$ 
corrections have been computed only for empty states that are close 
to the conduction band edge \cite{Chang00,Olevano02}, thus we can not
verify the accuracy of a scissor-operator approximation far from the
edge.

Recently Mo and Ching \cite{Mo01} have calculated isotropic Si 
and O $K$-XANES and the corresponding LDOS using a supercell 
approach within an {\it ab initio} orthogonalized linear 
combination of atomic orbitals (OLCAO) method. A good agreement 
has been obtained between experiments and XANES calculations. 
With regards to LDOS calculations, the situation is quite different. 
They have shown that the Si $K$-edge can be reasonably represented 
by Si $p$ LDOS, while O $p$ LDOS surprisingly fails to reproduce 
any feature of the experimental spectrum. This result would mean 
that the radial dipole matrix element are strongly energy dependent
in that case. This last consideration may be taken cautiously since the
energy dependence of the dipole matrix element as well as the LDOS
are widely dependent on the method used for their calculation. 
In particular, the LDOS depend on the size of the integration region
used for the angular momentum projection. Only if this region
is sufficiently small, the LDOS are independent of the
size of the region, appart from a multiplicative factor.
We calculated Si and O $p$ LDOS, using an integration sphere of
0.4~\AA\ radius. The resulting LDOS are similar to
the corresponding total cross sections computed with our approach,
in contradiction with the conclusions of Mo and Ching \cite{Mo01}.
To explain this discrepancy, it should be noticed that
Mo and Ching used the Mulliken population to compute the LDOS,
which corresponds to a much larger effective integration region.

\section{conclusion}

We have presented an {\it ab initio} framework based on pseudopotentials
for calculating XANES spectra. The scheme uses self-consistent 
full-potentials.  It was applied to the calculation of $K$-edge
spectra in diamond and $\alpha$-quartz. The results were compared with 
isotropic (diamond) and dichroic ($\alpha$-quartz) experimental data 
and a good agreement was obtained. In the case of diamond, our 
250 atom supercell calculated spectrum  is comparable with the BS 
calculated one of Soininen and Shirley. These first applications
have shown that the introduction of the recursion method permits the 
treatment of large supercells. This opens new fields of applications
like surfaces, small agregates or amorphous materials, for which large 
supercells are required. The continued fraction may also permit the 
investigation of more-extended x-ray absorption regions, without 
adding significant computation time. The  results suggest that our 
scheme is quite robust and could be further successfully applied in 
XANES and ELNES spectroscopies.  \\

We would like to thank G. Tourillon and C. Laffon for having provided 
polarized O $K$-edge experimental spectra. We also thank J.J. Rehr
for encouraging and fruitful discussions. The english was polished by
S. Joyce.  We are grateful to Ch. Brouder for his thorough reading of 
the manuscript. This is an IPGP contribution \#XXXX.

\newpage

\begin{thebibliography}{55}
\bibitem{footnote}
\bibinfo{note}{A natural choice for the {\it all electron} partial waves is the
  solutions of the radial Schr\"odinger equation for the isolated atom.}
\end{thebibliography}
\end{document}